\documentclass[twocolumn]{aastex631}

\usepackage{natbib}
\usepackage{hyperref}
\usepackage{appendix}

\begin{document}

\title{Explaining the Weak Evolution of the High-Redshift Mass-Metallicity Relation \\ with Galaxy Burst Cycles}

\author[0000-0001-6676-4132]{Andrew Marszewski\footnotemark}
\affiliation{CIERA and Department of Physics and Astronomy, Northwestern University, Evanston, IL 60201}

\author[0000-0002-4900-6628]{Claude-André Faucher-Giguère}
\affiliation{CIERA and Department of Physics and Astronomy, Northwestern University, Evanston, IL 60201}

\author[0000-0002-1109-1919]{Robert Feldmann}
\affiliation{Department of Astrophysics, University of Zurich, Zurich CH-8057, Switzerland}

\author[0000-0003-4070-497X]{Guochao Sun}
\affiliation{CIERA and Department of Physics and Astronomy, Northwestern University, Evanston, IL 60201}

\begin{abstract}
\footnotetext{Corresponding Author: Andrew Marszewski \\ \href{mailto:AndrewMarszewski2029@u.northwestern.edu}{AndrewMarszewski2029@u.northwestern.edu}}

Recent observations suggest a nearly constant gas-phase mass-metallicity relation (MZR) at $z \gtrsim 5$, in agreement with many theoretical predictions. This lack of evolution contrasts with observations at $z \lesssim 3$, which find an increasing normalization of the MZR with decreasing redshift. We analyze a high-redshift suite of FIRE-2 cosmological zoom-in simulations to identify the physical drivers of the MZR. Previous studies have explained the weak evolution of the high-redshift MZR in terms of weakly evolving or saturated gas fractions, but we find this alone does not explain the evolution in FIRE-2. Instead, stellar feedback following intense bursts of star formation drives enriched gas out of galaxies, resetting their interstellar medium and separating their histories into distinct ``burst cycles".  We develop the ``Reduced Burst Model", a simplified gas-regulator model that successfully reproduces the simulated MZR and identifies the dominant drivers of its evolution.  As redshift decreases, the metallicity of inflows within burst cycles increases at fixed stellar mass due to increased wind recycling of enriched gas.  Meanwhile, the metal mass produced by stars per inflowing gas mass within these cycles decreases because of decreased star formation per gas mass inflowing into the galaxy.  The effects of these two processes on the median metallicity largely cancel, holding the MZR constant for $z = 5 - 12$. At fixed stellar mass, the simulations predict lower gas metallicities at higher $\rm H\alpha$-derived star formation rates, in qualitative agreement with the fundamental metallicity relation (FMR), but this effect is reduced in rest UV-selected samples.

\end{abstract}

\section{Introduction} \label{sec:intro}

\begin{figure*}[t!]
    \centering
    \includegraphics[width=\linewidth]{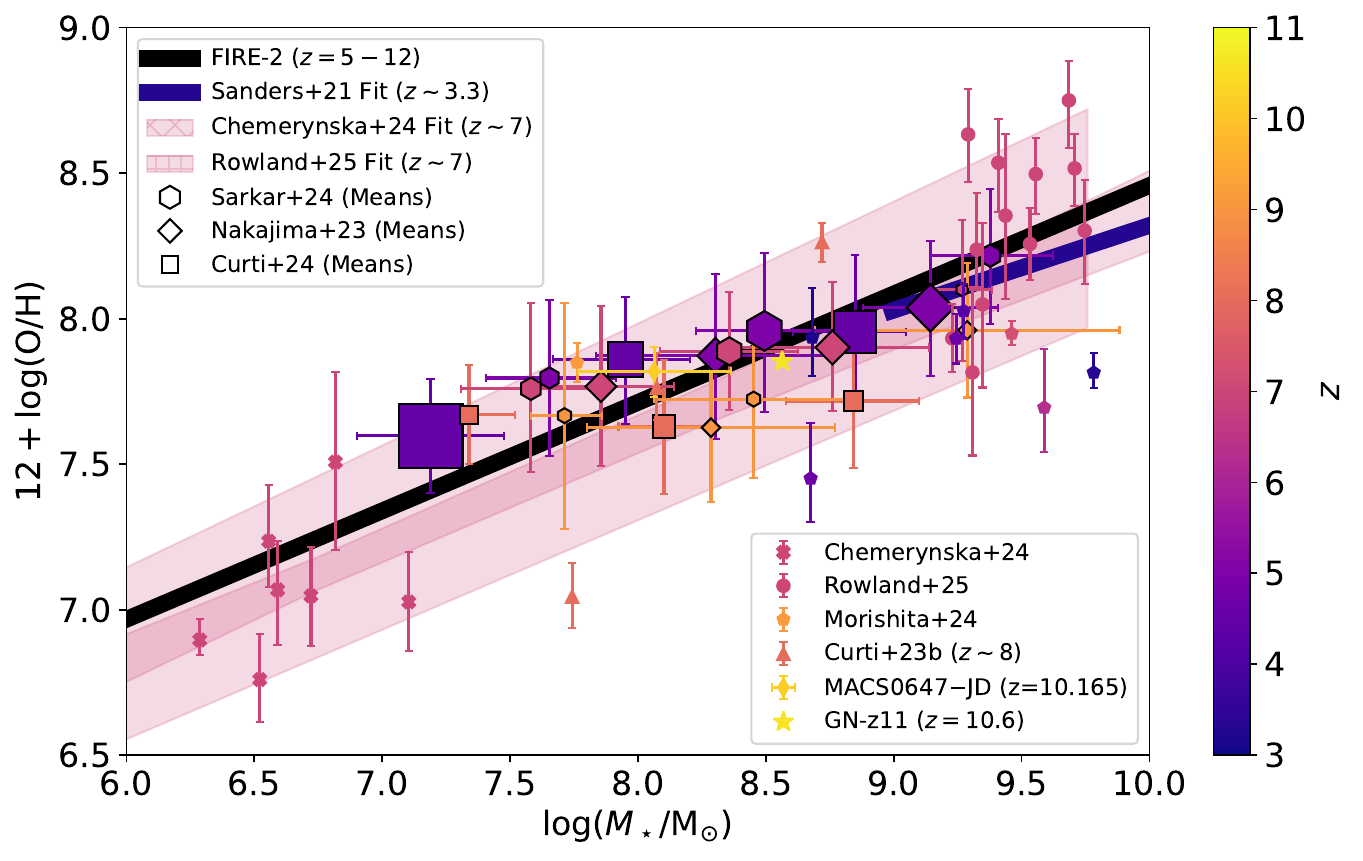}
    \caption{The high-redshift gas-phase MZR in FIRE-2 for $z=5-12$ from \citet{Marszewski_2024} (black solid line) and in observations from $z= 3-11$.  All observational data (fits, binned means, and individual galaxies) are colored by redshift.  Observational best fits are shown at $z \sim 3.3$ from \citet{Sanders_2021} (solid line) and at $z \sim 7$ from \citet{Chemerynska_2024} (diagonal hatching representing uncertainty in the best linear fit) and \citet{Rowland_2025} (orthogonal hatching representing the typical scatter of galaxies about the best fit). Shapes with black borders show stellar mass and redshift-binned means from \citet{Sarkar_2024} (hexagons), \citet{Nakajima_2023} (diamonds), and \citet{Curti_2023b} (squares).  The sizes (areas) of these shapes are proportional to the number of galaxies within these bins.  Shapes without black borders are measurements of individual galaxies from \citet{Chemerynska_2024} (X-markers), \citet{Rowland_2025} (circles), \citet{Morishita_2024} (pentagons), \citet{Curti_2023a} (triangles), \citet{Hsiao_2024} (MACS0647-JD; diamond), and \citet{Bunker_2023} (GN-z11; star).  The slope and normalization of the FIRE-2 MZR are in good agreement with high-redshift observations.  Moreover, the prediction of a weakly evolving MZR at high redshift is broadly consistent with JWST results which show little evidence for evolution from $z\sim 3-8$.  Some preliminary evidence for a decreased normalization in the MZR is measured at $z \gtrsim 8$, however, the galaxy samples in these high-redshift bins remain small and potentially have a stronger selection bias toward galaxies with higher recent star formation rates and therefore lower gas-phase metallicities given the existence of an FMR.}
    \label{fig:Observational_MZR}
\end{figure*}

The gas-phase mass-metallicity relation (MZR) is the observed positive correlation between a galaxy’s stellar mass and its gas-phase metallicity (\citealp{1979A&A....80..155L, Tremonti_2004}).  There is also an observed relation between a galaxy's stellar mass and its stellar metallicity (the stellar MZR), but throughout this work, we focus on the gas-phase MZR.  The MZR and its evolution have been observed extensively across wide ranges of redshift and stellar mass (e.g., \citealp{Erb_2006, Lee_2006, Zahid_2011, Zahid_2012, Henry_2013a, Henry_2013b, Maier_2014, Steidel_2014, Yabe_2014, Sanders_2015, Guo_2016}). \citet{Zahid_2013} characterized the observed evolution of the MZR from $z=0-2.3$, noting that, for a given stellar mass, metallicity tends to increase as redshift decreases.  However, observations are preliminarily consistent with a more weakly evolving MZR for $z \gtrsim 3$, as predicted by most theoretical models.  Figure \ref{fig:Observational_MZR} presents MZR observations for $z=3-11$.  The MZR's sensitivity to the cosmic baryon cycle processes that drive galaxy formation and evolution motivates an investigation into the physical drivers of its weakened redshift evolution.

The unprecedented spectroscopic capabilities of the James Webb Space Telescope (JWST) have expanded the redshift and stellar mass regimes where the MZR and its evolution have been measured.  \citet{Nakajima_2023} characterize the evolution of the MZR for $4<z<10$ using metallicity measurements of 135 galaxies identified by JWST in this redshift range.  They find no evidence beyond their level of error for evolution in the MZR for $z=4-10$.  \citet{Curti_2023b} analyze the gas-phase metallicities of 146 high-redshift ($3<z<10$) galaxies observed by JWST, 80 of which were also present in the sample from \cite{Nakajima_2023}.  \citet{Sarkar_2024} present the MZR for an additional sample of 81 star-forming galaxies with redshifts $4<z<10$.  They report some decrease in normalization of the MZR with increasing redshift. 
 However, this reported evolution is driven primarily by measurements in their highest-redshift ($z=8-10$) bins, which are sparsely sampled and potentially susceptible to selection bias effects that would result in a higher sampling rate of metal-poor galaxies at higher redshifts.  \citet{Bunker_2023} use strong line ratios to constrain the metallicity of GN-z11 at $z \sim 10.6$.  Gas-phase metallicities have been derived for a number of other high-redshift JWST targets via the direct $T_e$-based method (e.g., MACS0647-JD at $z = 10.165$; \citealp{Hsiao_2024}, galaxies in JWST Early Release Observations at $z \sim 8$; \citealp{Curti_2023a}, and 9 sources in the sight line of MACS J1149.5+2223 at $z = 3-9$; \citealp{Morishita_2024}).  \citet{Chemerynska_2024} and \citet{Rowland_2025} probe the very low-mass end and the very high-mass end of the MZR, respectively, at $z \sim 7$.

The MZR and its evolution have been studied at high redshift in a number of different simulation codes, such as IllustrisTNG \citep{Torrey_2019}, FirstLight \citep{Langan_2020}, SERRA \citep{Pallottini_2022}, ASTRAEUS \citep{Ucci_2023}, and FLARES \citep{Wilkins_2023}.  FirstLight, ASTRAEUS, and FLARES predict weak or no evolution in the MZR for $z \gtrsim 5$.  The Feedback in Realistic Environment (FIRE) project\footnote{See the FIRE project website: \url{http://fire.northwestern.edu}.} is a set of cosmological zoom-in simulations that resolve the multiphase ISM of galaxies and implement detailed models for star formation and stellar feedback (\citealp{Hopkins_2014,Hopkins_2018, Hopkins_2023}).  \citet{Ma_2016} characterized the MZR in the first generation of FIRE simulations from $z=0$ to $z=6$.  \citet{Marszewski_2024} analyzed the MZR in a high-redshift suite of FIRE-2 simulations and found that the MZR was in place at $z=12$ and held approximately constant down to $z=5$.  Figure \ref{fig:Observational_MZR} shows good general agreement between the form and weak evolution of the FIRE-2 MZR for $z=5-12$ and JWST observations from $z=3-11$.  A number of previous works (e.g., \citealp{Ma_2016, Torrey_2019, Langan_2020}) invoke gas fractions to explain the redshift evolution or lack of redshift evolution in the MZR.  Recently, \citet{Bassini_2024} analyzed the evolution of the MZR for $z=0-3$ in FIREbox \citep{Feldmann_2023}, a cosmological volume simulation that uses FIRE-2 physics, and found that evolving gas fractions are not responsible for the evolution in the MZR over this redshift range.  Rather, a combination of evolving inflow metallicities, outflow metallicities, and mass-loading factors drive the decrease in normalization of the MZR from $z = 0$ to $z = 3$. 

In this work, we connect the form and evolution of the MZR to the cosmic baron cycle processes (i.e., inflows, outflows, star formation, and stellar feedback) driving galaxy formation and evolution at high redshift in FIRE-2.   We find that strong feedback following bursty star formation drives massive outflows that reset the ISM of galaxies and separate their histories into distinct burst cycles.  We develop the ``Reduced Burst Model" for predicting the gas-phase metallicity of galaxies by identifying the cosmic baryon cycle processes that drive metallicity in the bursty, high-redshift regime.  Through explicit particle and galaxy tracking, we show that the ``Reduced Burst Model" successfully reproduces the form and weak redshift evolution of the MZR measured in FIRE-2 by \citet{Marszewski_2024} for $z=5-12$.   The scaling of the dominant cosmic baryon cycle properties identified in the ``Reduced Burst Model" with stellar mass are shown to reproduce the measured slope and normalization of the high-redshift MZR in FIRE-2.  Importantly, we identify the evolving processes that are responsible for the constancy of the MZR in FIRE-2 from $z=5-12$.  Evolution of the inflow metallicity and the metal production efficiency (the metal mass formed in stars per inflowing gas mass) largely balance one another, holding the MZR approximately constant.  We stress that the applicability of our ``Reduced Burst Model" is not dependent on the measured weak evolution of the MZR but only on the bursty formation histories of galaxies in this regime, so it can potentially be used to gain insight into physical regimes or other simulations where the MZR evolves more significantly.

We also investigate the existence of a dependence of gas-phase metallicity on star formation rate (SFR) in addition to stellar mass, as suggested by the fundamental metallicity relation (FMR), in our simulations.  The FMR states that, at fixed stellar mass, galaxies' metallicities are inversely correlated with their star formation rates (e.g., \citealp{Ellison_2008, Mannucci_2010}) or their gas content (e.g., \citealp{Bothwell_2013}).  We find evidence for a secondary dependence of metallicity on the $\rm H\alpha$-derived SFR at fixed stellar mass.  However, this signal weakens significantly when considering only galaxies that would be detected in rest-UV-selected surveys (which preferentially select galaxies that have had significant recent star formation), as is commonly the case for JWST observations. The secondary dependence on SFR appears to vanish when the SFR is inferred based on the continuum UV, which probes longer time scales.

This paper is organized as follows.  In Section \ref{sec:methods}, we describe the high-$z$ suite of FIRE-2 simulations used in this paper and the separation of these simulated galaxy histories into burst cycles for our analysis.  In Section \ref{sec:develop_analytic_model}, we develop an analytic “Reduced Burst Model" that identifies the physical processes that set the metallicities of galaxies in the bursty, high-redshift regime.  In Section \ref{sec:results}, we perform and present the results of our analysis on the “Reduced Burst Model" and its ability to identify the physical drivers of the weak evolution of the high-redshift MZR in FIRE-2.  In Section \ref{sec:discussion} we discuss in greater detail the physical processes and quantities that set the form and evolution of the MZR at high redshift.  We discuss previous, incomplete explanations of the weak evolution in the high-redshift MZR.  We contrast the scenario of weak evolution in the high-redshift MZR with the lower-redshift scenario, where the MZR is observed to evolve upward in cosmic time. 
We additionally discuss our FMR results in this section.  We summarize our key conclusions and future directions in Section \ref{sec:conclusions}.

Throughout this work, we adopt a standard flat $\Lambda$CDM cosmology with cosmological parameters consistent with \citet{Planck_2020}.  We adopt the solar metallicity, $Z_\odot = 0.02$, from \citet{Anders_1989}.

\begin{figure*}[t!]
    \centering
    \includegraphics[width=\linewidth]{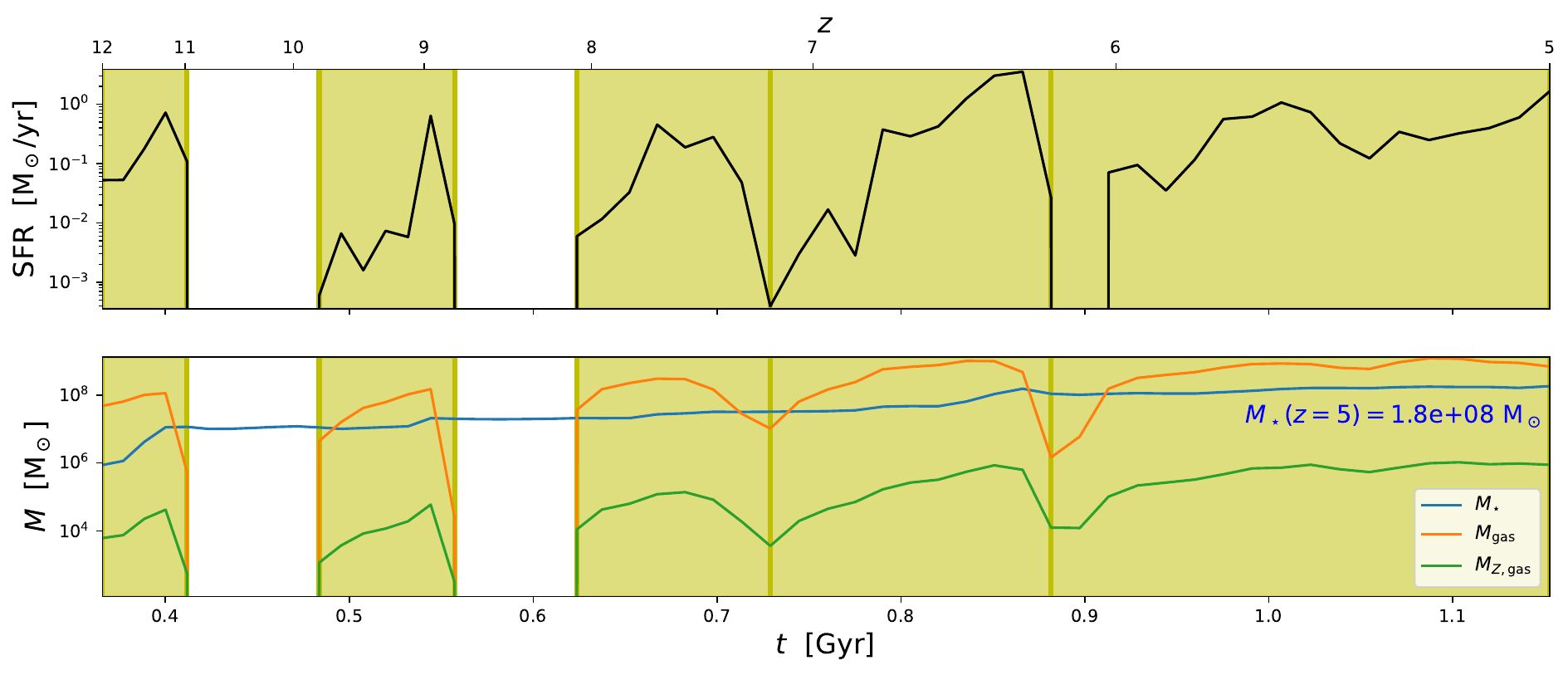}
    \caption{Star formation rate (upper) and stellar/gas/metal mass (lower) time series for an example high-redshift FIRE-2 galaxy with a stellar mass of $M_\star = 1.8 \times10^{8} \rm M_\odot$ at $z=5$.  The star formation rate (black) is characterized by strong bursts.  The stellar mass (blue) grows over time via these bursts.  Intense stellar feedback following the starbursts drives strong outflows, decimating the gas mass (orange) and gas-phase metal mass (green) within the galaxy.  We analyze individual burst cycles denoted here by highlighted regions separated by darker yellow lines.}
    \label{fig:Time_Series}
\end{figure*}

\section{Simulations and Galactic Burst Cycles} \label{sec:methods}
\subsection{The Simulations}

We utilize a high-redshift suite of FIRE-2 cosmological zoom-in simulations originally presented by \citet{Ma_2018a,Ma_2018b,Ma_2019}.  Notably, \citet{Sun_2023b} demonstrate that this simulation suite reproduces the high-redshift UV luminosity function (UVLF) measured by JWST.  Furthermore, the same FIRE-2 physics model also provides a good match to both the observed UVLF and UV luminosity density over $z=6-14$ \citep{Feldmann_2025}.  This high-redshift suite of FIRE-2 simulations was used by \citet{Marszewski_2024} to show that the MZR in FIRE-2 evolves very weakly for $z \geq 5$ and is in excellent agreement with JWST measurements.  This simulation suite was run using the GIZMO code \citep{Hopkins_2015}.  The hydrodynamic equations are solved using GIZMO's meshless finite-mass (MFM) method.  The 34 particular simulations analyzed in this paper are the z5m12a--e, z5m11a--i, z5m10a--f, z5m09a--b, z7m12a--c, z7m11a--c, z9m12a, and z9m11a--e runs.  The names of these simulations denote the final redshift that they were run down to ($z_{\rm fin}=5$, $7$, or $9$) and the main halo masses (ranging from $M_{\rm halo} \approx 10^9-10^{12} \rm{M}_\odot$) at these final redshifts.  Baryonic (gas and star) particles have initial masses $m_b = 100-7000 \rm{M}_\odot$ (simulations with more massive host galaxies have more massive baryonic particles).  Dark matter particles are more massive by a factor of $\Omega_{\rm DM}/\Omega_{\rm b} \approx 5$. Gravitational softenings are adaptive for the gas (with minimum Plummer-equivalent force softening lengths $\epsilon_b = 0.14-0.42$ physical pc) and are fixed to $\epsilon_* = 0.7-2.1$ physical pc and $\epsilon_{\rm DM} = 10-42$ physical pc for star and dark matter particles, respectively.

A full description of the baryonic physics in FIRE-2 simulations is given by \citet{Hopkins_2018}, while more details on this specific suite of FIRE-2 simulations are discussed in \citet{Ma_2018a,Ma_2018b,Ma_2019}.  Here, we briefly review the aspects of the simulations most pertinent to our MZR analysis. 

FIRE-2 simulations track the abundances of 11 different elements (H, He, C, N, O, Ne, Mg, Si, S, Ca, Fe).  Metals are returned via multiple stellar feedback processes, including core-collapse and Type Ia supernovae as well as winds from O/B and asymptotic giant branch (AGB) stars.  Star particles in the simulations represent stellar populations with a Kroupa initial mass function \citep{Kroupa_2001} and with the stellar evolution models from STARBURST99 \citep{Leitherer_1999}.  These simulations use the UV background model from \citet{CAFG_2009}.  These simulations also include subgrid modeling for turbulent diffusion of metals to allow for chemical exchange between neighboring particles.  The implementation and effects of the subgrid turbulent diffusion model in FIRE simulations are described in \citet{Colbrook_2017} and \citet{Escala_2018}.

Consistent with \citet{Marszewski_2024}, we define a galaxy's gas-phase metallicity to be the mass-weighted mean metallicity of all gas particles within $0.2R_{\rm vir}$ of the galaxy's center.  While this quantity is not directly measured in observations, \citet{Marszewski_2024} compare this definition of metallicity to others (i.e., SFR-weighted mean metallicity, alternative radial apertures, temperature cuts for identifying HII regions) and find that the resulting MZR is relatively insensitive to these different definitions.

\begin{figure*}[t!]
    \centering
    \includegraphics[width=\linewidth]{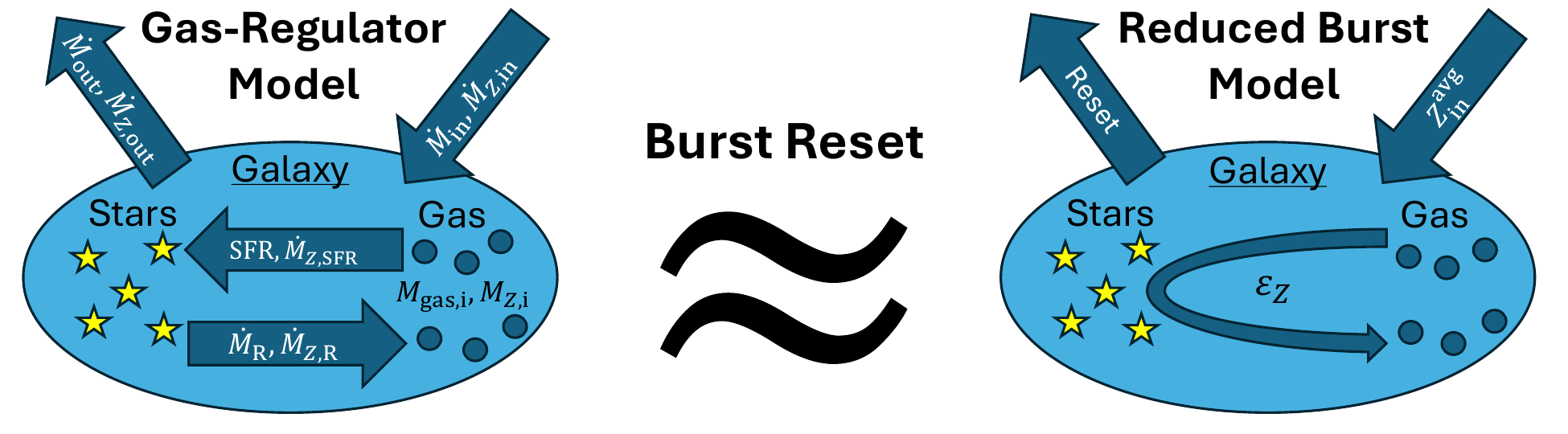}
    \caption{Channels by which galaxies can gain and lose metals and gas in the full “Gas-Regulator Model" (left; equation \ref{eqn:full_eqn}) and the “Reduced Burst Model" (right; equation \ref{eqn:simp_eqn}).  The full “Gas-Regulator Model" includes all possible channels (i.e., inflows, outflows, star formation, and stellar mass return).  The “Reduced Burst Model" includes only the channels relevant for explaining the form and evolution of the MZR in the scenario where galaxies have their ISM reset by bursty feedback.}
    \label{fig:Cartoon}
\end{figure*}

\subsection{Galactic Burst Cycles} \label{sec:burst cycles}

Figure \ref{fig:Time_Series} shows the star formation, stellar mass, gas mass, and gas-phase metal mass histories of an example high-redshift FIRE-2 galaxy with a stellar mass of $M_\star = 1.8 \times10^{8} \rm M_\odot$ at $z=5$ that undergoes five successive “burst cycles".  The histories of high-redshift galaxies in FIRE-2 are characterized by a series of “burst cycles", each of which typically contains the following three phases: (1) The galaxy accretes large amounts of relatively metal-poor gas, typically a mixture of pristine cosmic inflows and recycled gas from the circumgalactic medium (CGM). (2) After sufficient gas accretion, galaxies undergo intense bursts of star formation, producing large numbers of stars on a short timescale.  (3) The bursty stellar feedback that follows this star formation rapidly enriches the ISM while also driving massive outflows that cause the majority, if not all, of the galaxy's ISM to be ejected.  At this stage, the galaxy's ISM has been effectively reset.  Once the intense feedback subsides, the galaxy may again begin to accrete gas, and the cycle repeats.

The mass ejection of a galaxy's ISM between burst cycles is keenly important for the analysis and results of this work.  This process prevents galaxies from retaining gas that would be enriched through multiple cycles of star formation, effectively resetting a galaxy's ISM and its enrichment.  

\section{Derivation of the Reduced Burst Model} \label{sec:develop_analytic_model}

Previous works have developed various versions of the “Gas-Regulator Model" (e.g., \citealp{Finlator_2008, Peeples_2011, Lilly_2013, Dayal_2013, Feldmann_2015}), within which the metallicities of galaxies are regulated by the cosmic baryon cycle processes (i.e., inflows, outflows, star formation, stellar feedback and returns).  Here, we begin with a statement of the full “Gas-Regulator Model" which explicitly demonstrates the dependence of metallicity on each of these processes.  We then simplify this expression to develop a reduced model that is applicable in the bursty, high-redshift regime in order to isolate the physical processes responsible for the form and evolution of the MZR at high redshift.  Figure \ref{fig:Cartoon} illustrates this simplification. 

The gas-phase metallicity of a galaxy is equal to its total gas-phase metal mass ($M_{Z, \rm gas}$) divided by its total gas mass ($M_{\rm gas}$):  
\begin{equation} \label{eqn:definitional_eqn}
    Z_{\rm gas} = \frac{M_{Z, \rm gas}}{M_{\rm gas}}.
\end{equation}
Accounting for all channels by which galaxies can gain and lose metals and gas (inflows, outflows, star formation, stellar returns), we can write the gas-phase metallicity in the full “Gas-Regulator Model" as
\begin{equation} \label{eqn:full_eqn}
    Z_{\rm gas} = \frac{M_{Z,\rm i}+\int (\dot{M}_{Z,\rm in} + \dot{M}_{Z,\rm R} - \dot{M}_{Z,\rm out} - \dot{M}_{Z,\rm SFR}) dt}{M_{\rm gas,i}+\int (\dot{M}_{\rm in} + \dot{M}_{\rm R} - \dot{M}_{\rm out} - \rm SFR)dt},
\end{equation}
where $M_{Z,\rm i}$ and $M_{\rm gas,i}$ are the initial metal mass and gas mass, respectively, $\dot{M}_{\rm in(out)}$ is the mass inflow (outflow) rate, $\dot{M}_{Z,\rm in(out)} = Z_{\rm in(out)}\dot{M}_{\rm in(out)}$ is the metal mass inflow (outflow) rate, $\dot{M}_{\rm R(Z,\rm R)}$ is the rate at which stars return mass (metals) into the ISM, $\rm SFR$ is the star formation rate, $\dot{M}_{Z,\rm SFR}$ is the rate at which metals become locked in stars through star formation.  We will now simplify this full “Gas-Regulator Model" to develop a new “Reduced Burst Model" that is applicable for the high-redshift, bursty regime.  We will argue that we can neglect the majority of the terms in equation \ref{eqn:full_eqn} to first order.  Later, in Section \ref{sec:results}, we demonstrate the validity of these approximations in our simulations. 

If we begin integrating the physical galaxy properties in equation \ref{eqn:full_eqn} immediately following a massive, feedback-driven outflow (i.e., at the start of a burst cycle denoted by the left boundaries of the highlighted regions in Figure \ref{fig:Time_Series}), $M_{Z,\rm i}$ and $M_{\rm gas,i}$ become negligible compared to other terms in equation \ref{eqn:full_eqn}.  We then have
\begin{equation}
    Z_{\rm gas} \approx \frac{\int (\dot{M}_{Z,\rm in} + \dot{M}_{Z,\rm R} - \dot{M}_{Z,\rm out} - \dot{M}_{Z,\rm SFR}) dt}{\int (\dot{M}_{\rm in} + \dot{M}_{\rm R} - \dot{M}_{\rm out} - \rm SFR) dt}.
\end{equation}
We measure the ISM gas and metal mass lost directly to star formation (astration) to be small compared to the contributions of other terms.  Moreover, if stars on average form with metallicity equal to the current gas-phase metallicity of the galaxy (i.e., $\int \dot{M}_{Z,\rm SFR} dt = Z_{\rm gas}\int \rm SFR \ dt$), the star formation terms in the numerator ($\int \dot{M}_{Z,\rm SFR} \ dt$) and denominator ($\rm SFR$) exactly cancel (removing gas with metallicity equal to the galaxy's current metallicity does not change the galaxy's metallicity).  Additionally, we neglect $\dot{M}_{\rm R}$ in the denominator since its magnitude is only a fraction of the already negligible contribution from $\rm SFR$. We then have
\begin{equation} \label{eqn:intermediate_eqn}
    Z_{\rm gas} \approx \frac{\int (\dot{M}_{Z,\rm in} + \dot{M}_{Z,\rm R} - \dot{M}_{Z,\rm out}) dt}{\int (\dot{M}_{\rm in} - \dot{M}_{\rm out}) dt}.
\end{equation}

We also find that the outflow terms have a negligible effect on the predicted MZR.  These terms usually only become nonnegligible near the end of a burst cycle when the massive outflows have already been launched.  Therefore, if the timescale over which the ISM is evacuated by strong feedback at the end of burst cycles is short compared to the full timescale of burst cycles, the outflow terms are only capable of significantly impacting our predicted metallicities for a small fraction of the time.  In Appendix \ref{appendix:Time Scales} we show that the evacuation timescale ($t_{\rm Evac}\sim 10-30 \ \rm Myr$) is indeed short compared to the full burst cycle timescale ($t_{\rm Burst}\sim 70-200 \ \rm Myr$) with a typical ratio of $t_{\rm Burst}/t_{\rm Evac} \sim 6$ throughout our redshift range.  As a result, while outflows play a pivotal role in resetting the ISM of galaxies, the integrated outflow gas and metal masses are negligibly small at most times within a burst cycle.  Additionally, we measure outflow metallicities to be near the current metallicities of the galaxies they are launched from (i.e., $\int \dot{M}_{\rm Z,\rm out} dt \approx Z_{\rm gas}\int \dot{M}_{\rm out} \ dt$).  This minimizes the ability of the outflow terms to drive change in a galaxy's metallicity (removing gas with metallicity equal to the galaxy's current metallicity does not change the galaxy's metallicity).  Removing the outflow terms, we then have
\begin{equation} \label{eqn:simp_eqn_terms}
    Z_{\rm gas} \approx \frac{\int (\dot{M}_{Z,\rm in} + \dot{M}_{Z,\rm R}) dt}{\int \dot{M}_{\rm in} \ dt}.
\end{equation}

\begin{figure*}[t!]
    \centering
    \includegraphics[width=\linewidth]{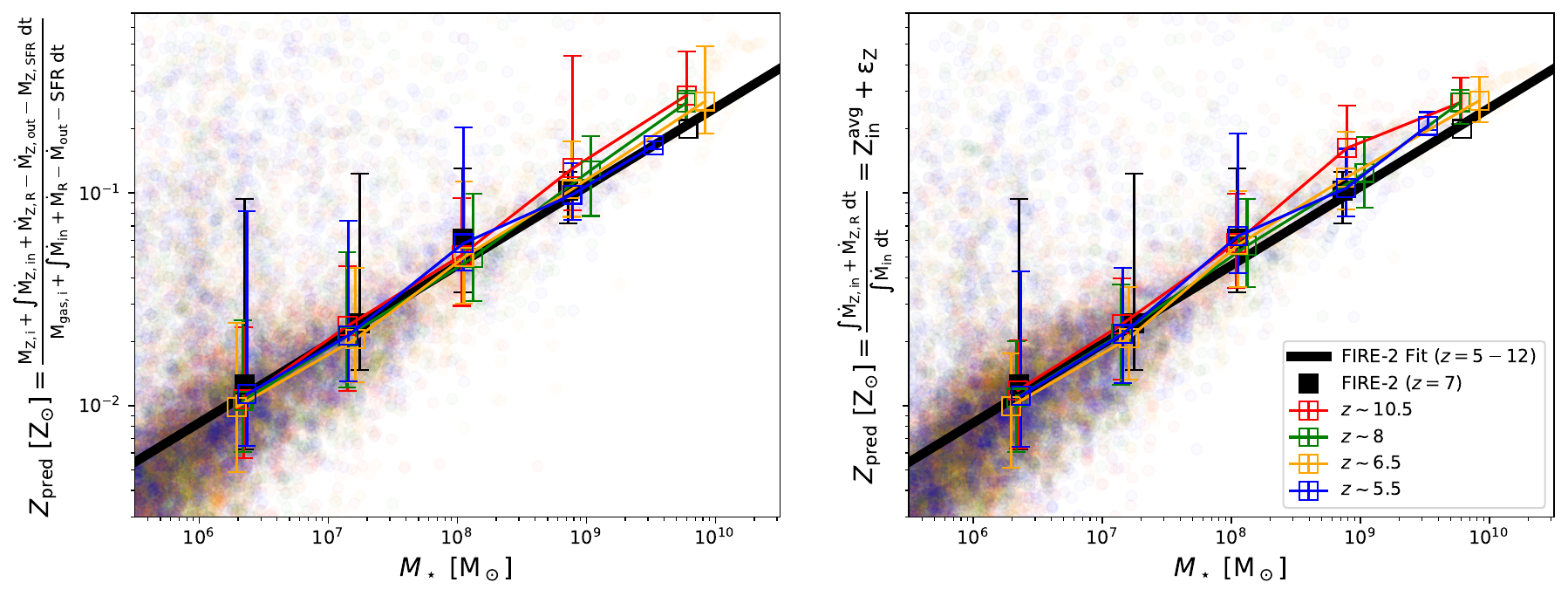}
    \caption{MZR of FIRE-2 galaxies predicted by the full “Gas-Regulator Model" (left; equation \ref{eqn:full_eqn}) and the “Reduced Burst Model" (right; equation \ref{eqn:simp_eqn}) at $z \sim 5.5$ (blue), $6.5$ (orange), $8.0$ (green), and $10.5$ (red).  Smaller, transparent points represent predicted metallicities of individual galaxies.  Empty Squares represent stellar mass-binned median predicted metallicities.  Black boxes show the MZR of FIRE-2 galaxies measured by \citet{Marszewski_2024} at $z=7$.  Error bars represent the 16th and 84th percentiles.  The black line shows the best-fit MZR reported by \citet{Marszewski_2024} for $z=5-12$ (no significant redshift evolution is predicted over this redshift interval).  Median metallicities predicted by both analytic models are in general agreement with the values measured directly from the simulations and match the prediction of a weakly evolving MZR at these redshifts.  The simplification between the full “Gas-Regulator Model" and the “Reduced Burst Model" has a negligible impact on the predicted MZR, motivating analysis of the reduced model to explain the form and evolution of the MZR for $z=5-12$.}
    \label{fig:MZR}
\end{figure*}

Finally, we define the time-averaged inflow metallicity within a burst cycle as $Z_{\rm in}^{\rm avg} = \int \dot{M}_{Z,\rm in} \ dt \ /\int \dot{M}_{\rm in} \ dt$ and the metal production efficiency as $\varepsilon_Z = \int \dot{M}_{Z,\rm R} \ dt \ /\int \dot{M}_{\rm in} \ dt$.  Our “Reduced Burst Model" is then
\begin{equation} \label{eqn:simp_eqn}
    Z_{\rm gas} \approx Z_{\rm in}^{\rm avg} + \varepsilon_Z.
\end{equation}
Figure \ref{fig:MZR} shows that metallicities calculated from the full “Gas-Regulator Model" (equation \ref{eqn:full_eqn}) are in general agreement with values measured directly from the simulations by \citet{Marszewski_2024}, matching the prediction of weak evolution in the MZR for $z \gtrsim 5$.  Moreover, these predictions do not appreciably change when using the “Reduced Burst Model" (equation \ref{eqn:simp_eqn}), validating our simplification from the full “Gas-Regulator Model" to the “Reduced Burst Model" above.  This motivates the use of our simplified model “Reduced Burst Model" to explain the weak evolution in the MZR that it predicts and that is measured in FIRE-2 at high redshift.

\section{Testing the Reduced Burst Model} \label{sec:results}

\subsection{Simulation Analysis}

We perform galaxy and particle tracking between snapshots from $z=5-12$ to extract the relevant quantities needed to calculate predicted metallicities via equations \ref{eqn:full_eqn} - \ref{eqn:simp_eqn} (i.e., $M_{\rm gas}$, $\rm SFR$, $\dot{M}_{\rm in}$, $\dot{M}_{\rm out}$, $Z_{\rm gas}$, $\dot{M}_{Z,\rm SFR}$, $Z_{\rm in}$, $Z_{\rm out}$).  Inflowing particles are identified as those that are within $0.2R_{\rm vir}$ of the galaxy center during the current snapshot but were not within that radius during the previous snapshot.  Similarly, outflowing particles are identified as those that are outside of this radius during the current snapshot but were within this radius during the previous snapshot.  To calculate stellar mass return rate ($\dot{M}_{\rm R}$) and the stellar metal production rate ($\dot{M}_{Z,\rm R}$), we plug the properties of stellar populations (mass, age, metallicity) into the FIRE-2 yields for O/B winds, AGB winds, Type Ia supernovae, and Type II supernovae provided in Appendix A of \citet{Hopkins_2018}.  

We measure a galaxy's gas-phase metallicity as the mass-weighted mean metallicity of all gas particles within $0.2 R_{\rm vir}$ and a galaxy's stellar mass as the total mass of all star particles within $0.2 R_{\rm vir}$.  In order to be tracked in our analysis, we require galaxies to have a nonzero stellar mass and a minimum virial mass given by $M_{\rm vir} \geq 10^9 M_\odot$.  We later enforce a stricter minimum stellar mass cutoff of $M_\star \geq 10^6 M_\odot$ for galaxies to be included in our final sample, matching the stellar mass cut used to characterize the MZR in \citet{Marszewski_2024}.  The initial cutoff to be tracked is less strict, since it can be important to begin tracking the properties of galaxies before they reach the more strict threshold to be included in our sample.  We exclude satellite galaxies and subhalos from our analysis, as their properties can be significantly influenced by their interaction with the halo of their host galaxy.

We divide the history of each tracked galaxy in our sample into a number of ``burst cycles".  A typical burst cycle begins with a galaxy accreting gas.  The accretion leads to intense, bursty star formation.  The bursts of stellar feedback following this star formation drive massive outflows. The galaxy remains gas-poor until it begins accreting gas again at the start of the next burst cycle.  We define the time boundaries of burst cycles in the following way using the gas mass histories of galaxies.  The start of the first burst cycle in a galaxy's history is the first time step for which the galaxy is tracked and has a minimum gas mass of $M_{\rm gas}^{\rm min} \geq 7000 \rm \, M_\odot$.  We have performed tests and found that the results of this work are insensitive to the choice in $M_{\rm gas}^{\rm min}$ over a large range ($M_{\rm gas}^{\rm min}=0-10^5 \, \rm M_\odot$).  The value of $7000 \rm \, M_\odot$ is chosen because it is the coarsest baryonic mass resolution in the simulation suite and is therefore the minimum gas mass that is resolvable for every simulation used. The end of a burst cycle is identified as the next time step for which the galaxy either has a gas mass below $M_{\rm gas}^{\rm min}$ or when its gas mass history reaches a local minimum that is below 50 percent of the current burst cycle's peak gas mass.  In the local minimum case, the final time step of the current burst cycle becomes the first time step of the next burst cycle.  In the case where the galaxy's gas mass reaches below $M_{\rm gas}^{\rm min}$, the next time step for which the galaxy has a gas mass larger than $M_{\rm gas}^{\rm min}$ becomes the first time step of the next burst cycle.  See Figure \ref{fig:Time_Series} for an example of a galaxy history divided into distinct burst cycles.

At each time step within a burst cycle, we evaluate the integrals in equations \ref{eqn:full_eqn} - \ref{eqn:simp_eqn} from the start of the current burst cycle to the current time step.  This method provides us with a predicted metallicity for each tracked galaxy at each time step for which the true gas-phase metallicity of the galaxy can also be directly measured with a resolved gas mass of at least $M_{\rm gas} \geq 7000 \, \rm M_\odot$.  To uncover the time evolution of the predicted MZR, we bin our calculated metallicities in stellar mass and cosmic time.  Metallicity predictions from each burst cycle are placed within one of four time bins according to the average cosmic time during that burst cycle.  These bins are centered at $z \sim 5.5, \ 6.5, \ 8.0, \ 10.5$.  Galaxies in each redshift bin are further separated into five stellar mass bins equally spaced in $\log M_\star$, matching the number of bins used in \citet{Marszewski_2024}.

In the next section, we present our term-by-term analysis of equation \ref{eqn:simp_eqn} across our redshift range ($z=5-12$).  We characterize the redshift evolution of galactic inflow properties and metal production, and we uncover how their effects on gas-phase metallicities balance one another, holding the MZR constant.  

\subsection{Performance and Implications of the Reduced Burst Model} 

\begin{figure*}[t!]
    \centering
    \includegraphics[width=\linewidth]{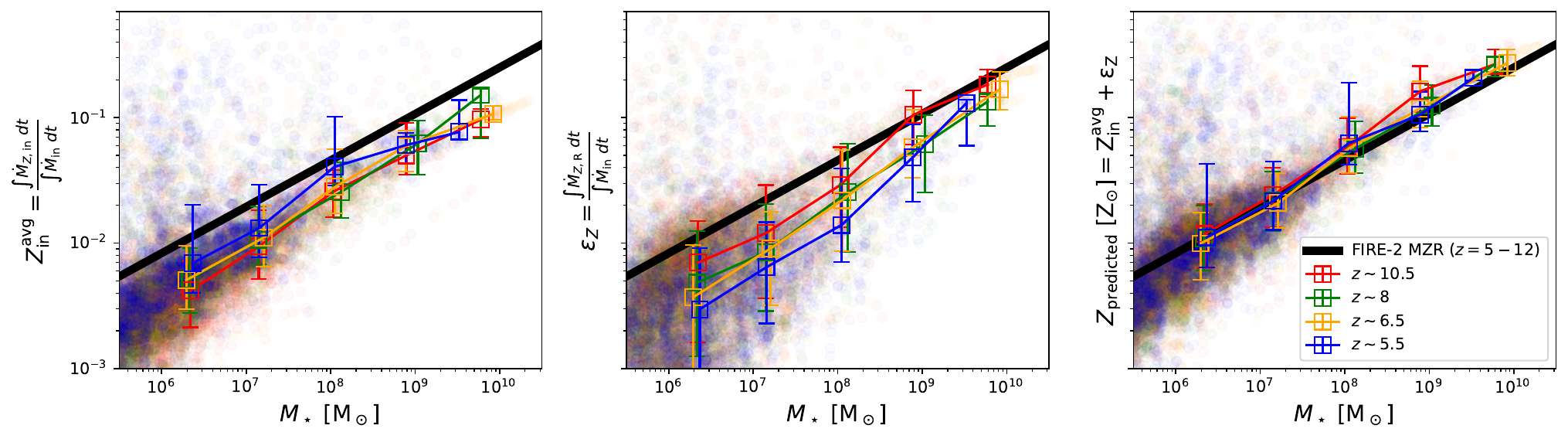}
    \caption{Average inflow metallicity $Z_{\rm in}^{\rm avg}$ (left), metal production efficiency $\varepsilon_Z$ (middle), and their sum (right) as a function of stellar mass at $z \sim 5.5$ (blue), $6.5$ (orange), $8.0$ (green), and $10.5$ (red).  Empty squares represent stellar-mass-binned median values.  Smaller, transparent points represent the measured quantities of individual galaxies at each time step within a burst cycle.  At fixed stellar mass, the average inflow metallicity decreases with increasing redshift, while the metal production efficiency increases.  The evolutions of these two quantities largely cancel one another out in their sum, resulting in the prediction of weak evolution in the MZR.}
    \label{fig:Term_Analysis}
\end{figure*}  

We present the performance and implications of our analysis on the “Reduced Burst Model" and its ability to identify the physical processes that drive the MZR at high redshift.  Figure \ref{fig:MZR} demonstrates the ability of both the full “Gas-Regulator Model" and the “Reduced Burst Model" to reproduce the form of the FIRE-2 MZR for $z=5-12$.  Moreover, both models successfully reproduce the prediction of a weakly evolving MZR at high redshift.  The success and simplified form of the “Reduced Burst Model" allow us to isolate the individual properties of the cosmic baryon cycle process that drive the form and constancy of the FIRE-2 MZR.

Figure \ref{fig:Term_Analysis} shows the redshift and stellar mass dependence of the average inflow metallicity $Z_{\rm in}^{\rm avg}$, the metal production efficiency $\epsilon_Z$, and their sum (the predicted gas-phase metallicity via the “Reduced Burst Model").  Notably, we find that the MZR is approximately constant for $z=5-12$ due to cancellation in the evolution of the average inflow metallicity ($Z_{\rm in}^{\rm avg}$) and the metal production efficiency $\epsilon_Z$ in equation \ref{eqn:simp_eqn}.  At fixed stellar mass, $Z_{\rm in}^{\rm avg}$ increases with decreasing redshift, while $\epsilon_Z$ decreases.  Their sum, shown in the rightmost panel in Figure \ref{fig:MZR} and Figure \ref{fig:Term_Analysis} to be an accurate predictor of a gas-phase metallicity, remains approximately constant over our redshift range.  Additionally, as discussed further in Section \ref{sec:slope} and shown in Appendix \ref{appendix:Scaling Relations}, the slope of the MZR may be understood through the power-law scaling relations of $Z_{\rm in}^{\rm avg}$ and $\epsilon_Z$ with stellar mass.

\section{Discussion} \label{sec:discussion} 

We now recap the physical quantities and processes that dictate the form and evolution of the MZR at high redshift.  We explore the differences between this high-redshift ($z\gtrsim5$) regime and the lower-redshift ($z \lesssim 3$) regime where the normalization is observed to evolve upward with decreasing redshift.  We discuss the shortcomings of previous explanations for the evolution of the high-redshift MZR. 
Finally, we investigate the dependence of gas-phase metallicity on the SFR at fixed stellar mass and redshift in the context of previous results on a possible fundamental metallicity relation.

\subsection{Slope of the High-z MZR} \label{sec:slope}

The origin of the slope of the high-redshift MZR can be understood from analysis of the scaling relations between the quantities in equation \ref{eqn:simp_eqn_terms} ($\int \dot{M}_{Z, \rm in} dt$, $\int \dot{M}_{Z, \rm R} dt$, and $\int \dot{M}_{\rm in} dt$) and stellar mass.  Figure \ref{fig:Scaling_Relations} in Appendix \ref{appendix:Scaling Relations} presents these scaling relations across our redshift and stellar mass range.  We measure that the integrated mass of inflowing gas scales with stellar mass as $\int\dot{M}_{\rm in}dt \propto M_\star^{0.75}$.  The scaling relations of $\int \dot{M}_{Z, \rm R} dt$ and $\int \dot{M}_{Z, \rm in} dt$ with stellar mass are slightly stronger than linear, with approximate scalings of $\int\dot{M}_{Z, \rm R}dt, \int\dot{M}_{Z, \rm in}dt\propto M_\star^{1.1}$.  The former scales slightly more strongly than the latter, but this approximate scaling holds well for each term throughout the stellar mass and redshift regime where the term in question contributes significantly to their sum.  As a result, the scaling relations of $\varepsilon_Z$ and $Z_{\rm in}^{\rm avg}$ with stellar mass both have power-law indices near the FIRE-2 high-redshift MZR slope of $0.37$ measured by \citet{Marszewski_2024}.  Although $\varepsilon_Z$ scales slightly more strongly and $Z_{\rm in}^{\rm avg}$ scales slightly more weakly with stellar mass, the scaling of their sum closely matches the scaling of the FIRE-2 MZR as shown in the right panel of Figure \ref{fig:Term_Analysis}.

\subsection{Normalization Evolution of the High-z MZR}

The massive outflows driven by the feedback following bursty star formation play a pivotal role in holding the MZR constant in our simulations at high redshift.  These outflows eject the majority, if not all, of the gas from galaxies, preventing large amounts of gas from being retained and enriched through multiple burst cycles.  If galaxies instead retained large amounts of gas throughout their histories, we would be unable to separate galaxies into distinct burst cycles that allow for the simplification between the full “Gas-Regulator Model" (equation \ref{eqn:full_eqn}) and our “Reduced Burst Model" (equation \ref{eqn:simp_eqn}).  It is only after noting the large amount of gas turnover between burst cycles that we are able to focus our analysis on the time-averaged inflow metallicity and metal production efficiency.

Upon the separation of galaxy histories into discrete burst cycles, we find that the MZR is approximately constant due to a cancellation between the evolution of the average inflow metallicity $Z_{\rm in}^{\rm avg}$ within a burst cycle and the evolution of the metal production efficiency $\varepsilon_Z$ within a burst cycle.  Here, we suggest explanations for the individual evolutions of $Z_{\rm in}^{\rm avg}$ and $\varepsilon_Z$. 

Inflow metallicity likely increases with decreasing redshift due to increased recycling of gas that was enriched and ejected in winds during previous burst cycles.  Inflowing gas consists of a mixture of pristine gas inflowing from the intergalactic medium and recycled gas from the CGM that was previously enriched within a galaxy, was ejected, and now reaccretes onto the galaxy.  The recycling of ejected gas and metals from the CGM to the ISM is a critical process of the cosmic baryon cycle and is characterized in FIRE simulations by \citet{Muratov_2017} and \citet{AA_2017} \citep[see also][]{Pandya_2021}.  As time goes on, more generations of gas are ejected from the galaxy's ISM, resulting in a higher portion of inflowing gas being recycled rather than pristine.

The metal production efficiency decreases with decreasing redshift as a result of a decrease in galaxy-scale star formation efficiency, defined here as the stellar mass formed per inflowing gas mass integrated within a burst cycle ($\rm SFE = \int \rm SFR \  dt /\int \dot{M}_{\rm in} \ dt$).  Note that this definition of $\rm SFE$ is distinct other commonly used star formation efficiencies, such as the halo-scale star formation efficiency ($\rm SFR / \dot{M}_{\rm halo}$, where $\dot{M}_{\rm halo}$ is the halo growth rate) and the stellar baryon fraction ($ M_\star/(f_bM_{\rm halo})$, where $f_b = \Omega_b/\Omega_m$) which are respectively shown by \citet{Feldmann_2025} and \citet{Ma_2018b} to be nonevolving in FIRE-2 at these high redshifts.  However, consistent with our findings here, \citet{Ma_2018b} find that, at fixed stellar mass and within our redshift range, galaxies at higher redshift have, on average, higher recent star formation rates than galaxies at lower redshift.  In our simulations, we measure a continued decrease in the star formation rate per inflowing gas mass within burst cycles at fixed stellar mass from $z=12$ to $z=5$ (see Appendix \ref{appendix:Star_Formation}).  This evolution in star formation efficiency may be explained by the denser ISMs of higher-redshift galaxies producing stars more rapidly prior to star formation being disrupted by stellar feedback.  Within a burst cycle, the mass of metals produced by stars is closely linked to the stellar mass formed because metal production is dominated by channels that closely follow star formation (i.e., core-collapse supernovae and O/B winds).  This results in less metal production from young, massive stars per inflowing gas mass.

\subsection{Differences with the Low-redshift Regime}
\label{sec:lowredshift}

The constancy of the MZR at $z>5$ found in FIRE-2 simulations is in contrast to the evolution of the MZR measured at lower redshift ($z \lesssim 3$) in both observations and theoretical models, including in the FIRE-2 model itself \citep{Bassini_2024}.  In this lower-redshift regime, the normalization of the MZR is observed to increase with decreasing redshift (i.e., a low-redshift galaxy is typically more enriched than a high-redshift galaxy of the same stellar mass).  While our analysis does not include this lower-redshift regime, we provide tentative explanations for the differences between galaxies in the high- and low-redshift regimes that result in this difference in evolution.

One major reason why the MZR evolves at lower redshift may be that galaxies begin to retain a significant amount of their gas between starbursts, invalidating the use of our “Reduced Burst Model" at lower redshifts.  We have performed tests (not shown here) demonstrating that galaxies at lower redshift in FIRE-2 typically retain larger amounts of gas between burst cycles than galaxies at higher redshift at fixed stellar mass.  This may be due to bursts in star formation becoming less intense at fixed stellar mass as redshift decreases.  The trend of decreasing star formation rate per inflowing gas mass within burst cycles at fixed stellar mass from $z=12$ to $z=5$ (see Appendix \ref{appendix:Star_Formation}) may continue down to lower redshift.  This would result in the weakening of the bursty feedback that follows, allowing galaxies to maintain a significant portion of their ISM between burst cycles.  ISM enrichment would then become a function of time that gas remains within the galaxy, which increases in cosmic time in the case that the ISM is not fully ejected between burst cycles.  A subtle manifestation of this effect may already be in place at the high stellar mass ($M_\star \gtrsim 10^9 \rm \, M_\odot$) end of our high-redshift sample.  Here, small deviations between the measured FIRE-2 MZR and the predictions of our simplified model (see the rightmost panel in Figure \ref{fig:Term_Analysis}) may be due to high-mass galaxies retaining some gas between burst cycles.

A second potential reason for the MZR beginning to evolve at lower redshift, applicable in regimes where our ``Reduced Burst Model" still holds, is that the evolutions in $Z_{\rm in}^{\rm avg}$ and $\varepsilon_Z$ may no longer balance one another to the same quantitative degree that we find at high redshift. 
This is actually expected, as the near-perfect cancellation between between $Z_{\rm in}^{\rm avg}$ and $\varepsilon_Z$ we find at high redshift appears to be somewhat fortuitous. 
A possible scenario is that $Z_{\rm in}^{\rm avg}$ continues to evolve upward as redshift decreases due to more generations of enriched gas being recycled at lower redshift.  Meanwhile, $\varepsilon_Z$ may stop decreasing with decreasing redshift as the stellar mass formed per inflowing mass stabilizes or even begins increasing due to outflows becoming weaker, causing a larger percentage of inflowed gas to be converted into stars rather than being ejected.

\subsection{Previous Analytic Explanations} \label{sec:Previous_Explanations}

Other simulations and seminumerical models have also predicted weak evolution in the high-redshift MZR (e.g., FirstLight; \citealp{Langan_2020}, ASTRAEUS; \citealp{Ucci_2023}, FLARES; \citealp{Wilkins_2023}, FIRE-1; \citealp{Ma_2016}), while IllustrisTNG predicts more pronounced evolution \citep{Torrey_2019}.  Some analytic models have been put forth to explain the form and evolution of the MZR (e.g., \citealp{Finlator_2008, Peeples_2011, Lilly_2013, Dayal_2013, Feldmann_2015}).  In particular, some previous works (including \citealp{Ma_2016, Torrey_2019, Langan_2020}) invoke gas fractions in these analytic models to explain the redshift evolution or lack of redshift evolution in the MZR.  Here, we expand on the explanation in \citet{Marszewski_2024} as to why gas fractions cannot be solely responsible for the weak evolution of the high-redshift MZR we observe in FIRE-2. 

\citet{Ma_2016} apply the “closed box" model and \citet{Langan_2020} utilize the “leaky box" model to describe metallicities in their simulations.  In the “closed box" model, galaxies (or halos) are described as a fixed collection of gas and stars without inflows and outflows.  Note that \citet{Ma_2016} apply the “closed box" model to entire halos rather than to galaxies since halos are less susceptible to the effects of inflows and outflows. Metallicity can then be written as,
\begin{equation} 
    Z_{\rm gas} = -y \ln({\tilde{f}_{\rm gas}}),
\end{equation}
where $y = M_{\rm Z}/M_\star$ (often assumed to be $y=0.02$) is the metal yield (the mass of metals returned to the ISM per unit mass in formed, long-lived stars) and $\tilde{f}_{\rm gas}$ is the version of the gas fraction given by $M_{\rm gas}/(M_\star + M_{\rm gas})$.  In the “leaky box" model,
\begin{equation} 
    Z_{\rm gas} = -y_{\rm eff} \ln({\tilde{f}_{\rm gas}}),
\end{equation}
where $y_{\rm eff}$ is the effective metal yield, which takes into account that some metals can “leak" out of the galaxy via outflows and is often calibrated to make the “leaky box" model best fit the data.  In these models, weak evolution of the MZR at high redshift is attributed to saturated and/or weakly evolving values of $\tilde{f}_{\rm gas}$. However, in reality, galaxies are not closed boxes, and metallicity is sensitive to the evolving properties of inflows and outflows in ways that are not explicitly captured by the effective yield in the “leaky box" model.  Moreover, \citet{Ma_2016} find that even the metallicities of halos are offset below the “closed box" prediction as a result of stars preferentially forming near the centers of halos, where metallicity is higher.  We therefore find the closed/leaky box explanation for the weak evolution in the high-redshift MZR to be, at best, incomplete, since it fails to explicitly capture the effect that evolving properties of inflows, outflows, and other processes have on metallicity through the effective yield. 

Equilibrium gas-regulator models (e.g., \citealp{Lilly_2013, Feldmann_2015}) predict the equilibrium metallicity of galaxies by quantifying the effects of various cosmic baryon cycle processes on their gas and metals.  \citet{Torrey_2019} explain the evolution in the high-redshift MZR via evolution in the gas fraction (defined as $f_{\rm gas}=M_{\rm gas}/M_\star$) within a version of the equilibrium gas-regulator model.  The regulator model gives an approximate equilibrium metallicity of the form (e.g., \citealp{Lilly_2013}),
\begin{equation} \label{eqn:grm}
    Z_{\rm eq} = Z_{\rm in} + \frac{y}{1+(1-R)^{-1}\eta+f_{\rm gas}},
\end{equation}
$Z_{\rm in}$ is the metallicity of accreted gas and $\eta = \dot{M}_{\rm wind} / \rm{SFR}$ is the mass-loading factor of galactic winds.  However, \citet{Bassini_2024} show by measuring the magnitude and evolution of each contributing term that the evolution of the gas fraction does not drive the evolution of the MZR in FIREbox, which uses the FIRE-2 code, at lower redshifts ($z=0-3$). This is because the magnitude of $f_{\rm gas}$ is generally small compared to other terms that appear in the denominator of Equation \ref{eqn:grm}.  Rather, evolution in the wind mass-loading factor and the metallicities of inflows and outflows at fixed stellar mass drive the decrease of the MZR with increasing redshift up to $z=3$.  While we present a different framework for metallicity in this work, similar analysis that we conducted from $z=5-12$ yielded results consistent with \citet{Bassini_2024}.  In particular, we find that evolution of the gas fraction as it manifests in Equation \ref{eqn:grm} would have a subdominant effect compared to the evolving properties of inflows, outflows, and metal production.

\subsection{The Dependence on SFR} \label{sec:FMR}

The fundamental metallicity relation (FMR) posits a three-dimensional relationship between a galaxy’s metallicity, its stellar mass, and its star formation rate (e.g., \citealp{Ellison_2008, Mannucci_2010}) (or its gas content; e.g., \citealp{Bothwell_2013}).  A common (but unproven) interpretation is that the accretion of more metal-poor gas results in a galaxy having both a lower observed metallicity and a higher observed star formation rate.  In the context of gas-regulator models, the secondary dependence of metallicity on gas content has previously been interpreted by the appearance of $f_{\rm gas}$ in the version of the gas-regulator model given by equation \ref{eqn:grm}.  Since, as discussed in Section \ref{sec:Previous_Explanations}, this $f_{\rm gas}$ dependence does not explain the evolution of the MZR in our simulations, it is natural to ask whether there is an FMR-like relation in our simulations.

Another question that follows naturally is whether such a relation is redshift-invariant (a “Strong FMR") or if metallicity has a continued dependence on a secondary parameter but in a way that evolves with redshift (a “Weak FMR").  The “Strong" vs “Weak" FMR terminology was previously used in \citet{Garcia_2024}.  The existence of a redshift-invariant (“Strong") FMR is an active area of debate, with some observational samples showing a continuation of the relation out to at least $z=2-3$ (e.g., \citealp{Sanders_2021}) and others finding much weaker evidence for an FMR at this redshift (e.g., \citealp{Nathalie_2025}).  JWST observations at $z \gtrsim 4$ have shown offsets between measured metallicities and predictions made by the FMR (e.g., \citealp{Langeroodi_2023, Nakajima_2023, Scholte_2025}), suggesting that the FMR may either evolve or not apply at very high redshift.  \citet{Garcia_2024} and \citet{Garcia_2025} find, with statistical significance, that a “Weak FMR" performs better than a “Strong FMR" in fitting metallicities from Illustris, IllustrisTNG, EAGLE, and Simba cosmological simulations.  Theoretically, there is no clear reason an FMR would be universal out to arbitrarily high redshifts.  We therefore focus here on the existence of a “Weak FMR" - a secondary dependence of metallicity on star formation rate that may evolve with redshift.

\begin{figure*}[t!]
    \centering
    \includegraphics[width=\linewidth]{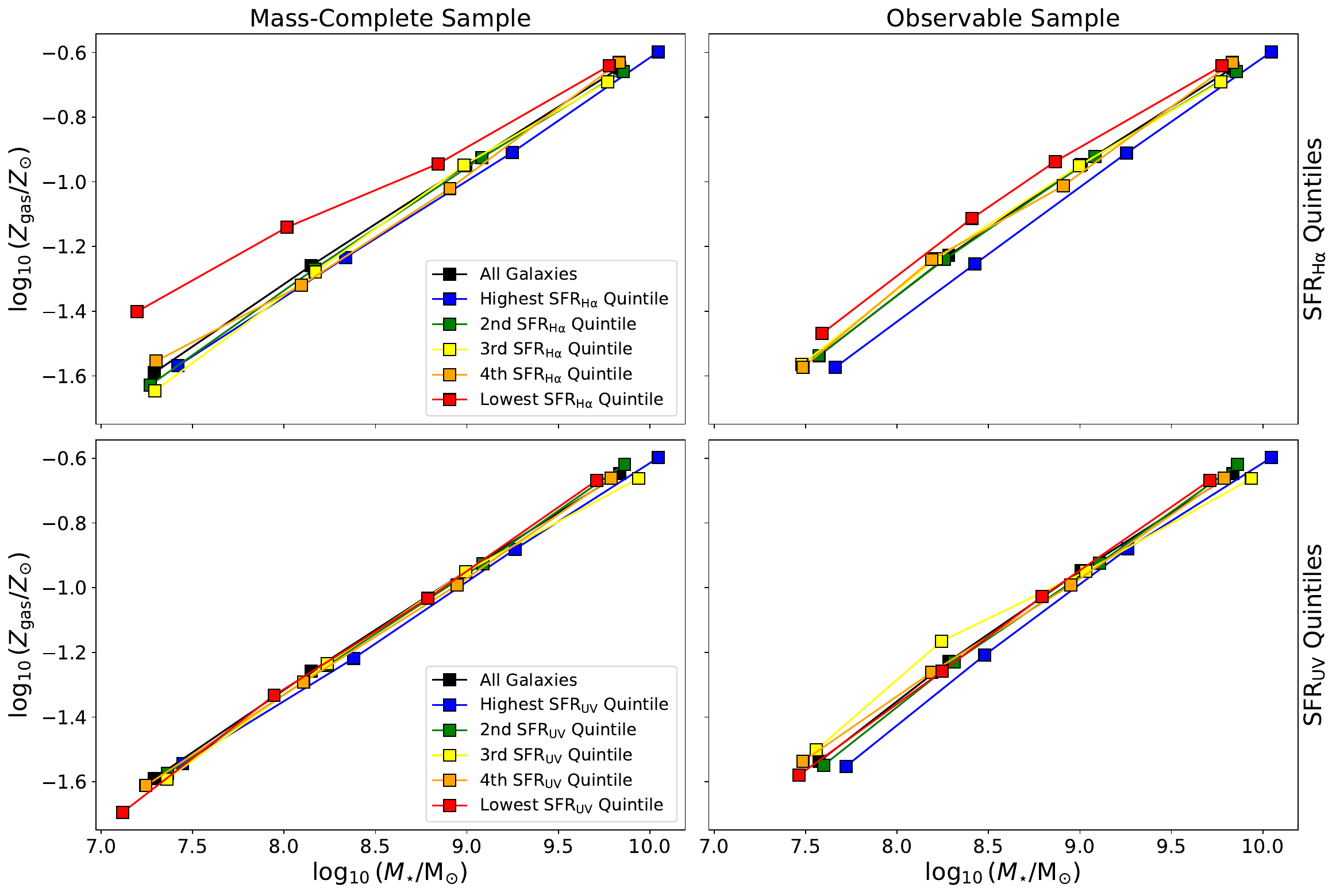}
    \caption{Signal for an FMR-like relation with $\rm SFR_{\rm H\alpha}$ (top) and $\rm SFR_{\rm UV}$ (bottom) as the secondary parameters for our complete sample of galaxies (left) and for an ``observable" ($L_{1560} > 2\times10^{43}$ erg/s, representative of a JADES-Deep-like survey) sample of galaxies (right).  Within each stellar mass bin, we plot the median metallicity of galaxies in the highest (blue), second-highest (green), third-highest (yellow), fourth-highest (orange), and lowest (red) SFR quintiles.  For our mass-complete sample, the lowest $\rm SFR_{\rm H\alpha}$ quintiles have elevated median metallicities, consistent with the existence of an FMR-like relation.  This signal weakens significantly when only considering the ``observable" galaxies in the sample.  There is little evidence for an FMR-like relation when using $\rm SFR_{\rm UV}$ as the secondary predictor for both the mass-complete and ``observable" sample.}
    \label{fig:FMR}
\end{figure*}

We investigate the potential existence of a secondary dependence of metallicity on star formation rate (an FMR-like relation) for galaxies in our simulations.  To do this, we divide galaxies in our sample into stellar mass bins centered between $M_\star \sim 10^7-10^{10} \, \rm M_\odot$.  Within each stellar mass bin, we further bin galaxies into quintiles (5 bins containing equal numbers of galaxies) according to their star formation rates.  We explore both UV-continuum and H$\rm \alpha$ emission as separate indicators for star formation rate.  We extract $\rm SFR_{UV}$, $\rm SFR_{H\alpha}$, and intrinsic UV luminosity by modeling the rest-frame UV/optical emission spectra of galaxies using BPASS v2.2 with nebular emission (lines and continuum) \citep{Stanway_2018}.  We then analyze the dependence of metallicity on star formation rate at fixed stellar mass. 

\citet{Sun_2023a} have previously demonstrated the observational selection effects that result from the rest-ultraviolet selection of galaxy samples at high redshift.  With this motivation, we investigate potential observational selection effects on an FMR-like signal by repeating the analysis described above but only including galaxies with intrinsic UV luminosities of $L_{1566} > 2\times10^{43}$ erg/s where $L_{1566} = \langle \lambda L_{\lambda} \rangle$ averaged over $\lambda=1556-1576$ \AA.  At $z=8$ (near the center of our redshift range), this cut approximately corresponds to a limiting magnitude of $m^{\rm lim}_{\rm AB}\sim30$, similar to the detection threshold for a JADES-Deep-like survey (e.g., \citealp{Robertson_2023}).

Figure \ref{fig:FMR} presents the signal found for an FMR-like relation in our simulations at $z = 5-12$.  For our full (mass-complete) sample of galaxies, the lowest $\rm SFR_{\rm H \alpha}$ quintile exhibits an elevated median metallicity compared with the other quintiles, consistent with the existence of an FMR-like relation.  This signal, however, is not seen when using $\rm SFR$ derived from UV-continuum emission.  We can understand the origin of the FMR-like signal from our “Reduced Burst Model" in equation \ref{eqn:simp_eqn_terms},
\begin{equation} \label{eqn:2nd_simp}
    Z_{\rm gas} = \frac{M_{Z, \rm gas}}{M_{\rm gas}} \approx \frac{\int (\dot{M}_{Z,\rm in} + \dot{M}_{Z,\rm R}) dt}{\int \dot{M}_{\rm in} \ dt}.
\end{equation}
This suggests that the secondary dependence of metallicity on $\rm SFR_{\rm H \alpha}$ is driven by metal-poor inflows of gas.  These inflows decrease a galaxy's metallicity while also providing the fuel for star formation, creating the inverse relation between $Z_{\rm gas}$ and $\rm SFR$ at fixed stellar mass.  This interpretation is also consistent with the stronger FMR signal found using $\rm SFR_{\rm H \alpha}$ as compared to that using $\rm SFR_{\rm UV}$.  $\rm H \alpha$ emission is sensitive to star formation on shorter timescales than the UV-continuum luminosity (e.g., \citealp{Kennicutt_2012, FV_2021}). As a result, we expect $\rm H \alpha$ emission to be more correlated with the presence of metal-poor gas that has fueled recent star formation than is UV luminosity which can persist following the enrichment of gas and the launching of outflows. 
In Appendix \ref{appendix:Time Scales} we show that the typical timescale for evacuation of the ISM at the end of a burst cycle is between $10-30 \ \rm Myr$.  This timescale is longer than the timescale traced by $\rm SFR_{\rm H \alpha}$ but comparable to or shorter than that traced by $\rm SFR_{\rm UV}$.

The overall signal for an FMR-like relation using $\rm SFR_{\rm H \alpha}$ becomes less clear when considering only the subsample of galaxies detectable by a JADES-Deep-like survey.  This is a result of our observability cut removing galaxies from our sample with little recent star formation that are typically more metal-rich.  While there is arguably still a signal for an FMR-like relation in our “observable" sample, we predict that observational selection effects will make it more difficult to detect such a relation in observed galaxy populations at high redshift.

\section{Conclusions} \label{sec:conclusions}
We have performed galaxy and particle tracking on high-redshift FIRE-2 galaxies to uncover the physical drivers of the form and evolution of the high-redshift MZR.  We have found that these galaxies' histories are characterized by distinct “burst cycles".  Within these cycles, metallicities can be accurately predicted via a new “Reduced Burst Model" analytic framework that only takes into account inflow properties and metal production in stars.  We have shown that cancellation in the evolution of average inflow metallicity ($Z_{\rm in}^{\rm avg}$) and metal production efficiency ($\varepsilon_Z$) hold the MZR constant at high redshift.  Here, we summarize the key conclusions of this work:
\begin{itemize}
  \item   
  Within a typical burst cycle,  galaxies first accrete gas.  This accretion leads to a burst of star formation.  The intense stellar feedback following this star formation drives massive outflows from the galaxy, effectively resetting its ISM. 
  \item Both a full ``Gas-Regulator Model" and our ``Reduced Burst Model'' are able to accurately predict the metallicities of high-redshift FIRE-2 galaxies.  The ``Reduced Burst Model" differs from the full ``Gas-Regulator Model" in that it only includes the baryon cycle processes that are dominant within burst cycles and primarily drive evolution in the MZR at high redshift.
  \item As redshift decreases from $z=12$ to $z=5$, we measure the average inflow metallicity $Z_{\rm in}^{\rm avg} = \int \dot{M}_{Z,\rm in} \ dt \ /\int \dot{M}_{\rm in} \ dt$ to increase and the metal production efficiency $\varepsilon_Z = \int \dot{M}_{Z,\rm R} \ dt \ /\int \dot{M}_{\rm in} \ dt$ to decrease.  The effects of the evolution of these two quantities on the MZR cancel one another for $z=5-12$, holding the MZR constant. 
  Previous explanations that focus on gas fractions to explain the (non)evolution of the high-redshift MZR do not adequately describe the evolution of FIRE-2 galaxies.
  \item We find evidence for an FMR-like secondary dependence of gas-phase metallicity on $\rm SFR_{\rm H\alpha}$.  However, the strength of this dependence is significantly reduced when considering only galaxies that are detected in surveys with rest-UV selections, as is common with JWST at high redshift.  We do not find significant evidence for such a relation when using the SFR derived from the UV-continuum luminosity, which probes longer timescales.  The secondary correlation with $\rm SFR_{\rm H\alpha}$ but not with $\rm SFR_{\rm UV}$ is consistent with the FMR being driven by the accretion of metal-poor gas that has fueled recent star formation.
\end{itemize}

Looking forward, the analysis of galactic burst cycles will be a powerful framework for connecting observed phenomena to the cosmic baryon cycle processes that drive galaxy formation and evolution. 
We note that while we used our ``Reduced Burst Model'' to explain the main factors producing a nearly constant MZR in the high-redshift limit in FIRE-2, the framework is applicable more generally.  It can be used to gain insight into other regimes or simulations where galaxies experience burst cycles, even if the MZR does evolve significantly. 

In future work, it would therefore be valuable to further explore the regimes where the framework is applicable, such as in the low-redshift but low-stellar-mass regime, where galaxies can also be bursty (e.g., \citealp{CAFG_2018}).  In particular, the framework could be used to better understand the physical processes driving observed chemical enrichment patterns (e.g., potential evolution in the FMR or the demographics of $\alpha$-to-Fe abundance ratios).  It would also be interesting develop the framework to better understand the drivers of other observed phenomena at high redshift connected to bursty star formation, such as extreme emission line galaxies (e.g., \citealp{Boyett_2024}) and mini-quenched galaxies (e.g., \citealp{Looser_2024}).

\section*{Acknowledgments}

The authors thank Nathalie Korhonen Cuestas and Allison Strom for useful discussions.  AM was supported by a CIERA Board of Visitors Fellowship.  GS was supported by a CIERA Postdoctoral Fellowship.  CAFG was supported by NSF through grants AST-2108230, AST-2307327, and CAREER award AST-1652522; by NASA through grants 17-ATP17-0067 and 21-ATP21-0036; by STScI through grants HST-GO-16730.016-A and JWST-AR-03252.001-A; and by CXO through grant TM2-23005X.  RF acknowledges financial support from the Swiss National Science Foundation (grant nos PP00P2-194814 and 200021-188552). The simulations analyzed in this work were run on XSEDE computational resources (allocations TG-AST120025, TG-AST130039, TG-AST140023, and TG-AST140064).  Analysis was done using the Quest computing cluster at Northwestern University.

\bibliography{bib}{}
\bibliographystyle{aasjournal}

\begin{appendices}

\section{Evolving Star Formation Efficiency} \label{appendix:Star_Formation}

\begin{figure*}[t!]
    \centering
    \includegraphics[width=\linewidth]{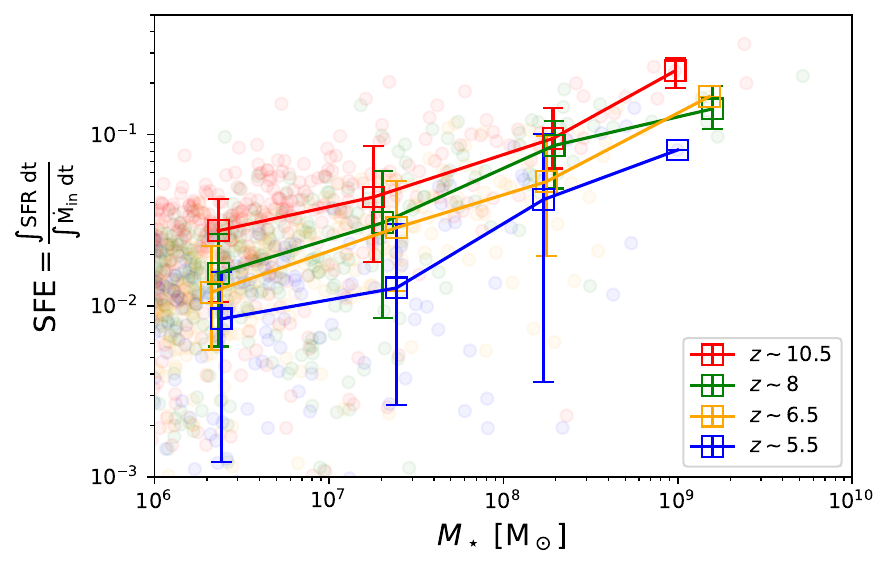}
    \caption{Star formation efficiency (defined here as $\rm SFE = \int \rm SFR \  dt /\int \dot{M}_{\rm in} \ dt$) as a function of stellar mass at $z \sim 5.5$ (blue), $6.5$ (orange), $8.0$ (green), and $10.5$ (red).  Smaller, transparent points represent the measured quantities integrated over individual galaxy burst cycles.  Empty Squares represent stellar-mass-binned median values.  At fixed stellar mass, the star formation efficiency increases with increasing redshift.  This increase results in the increase of $\varepsilon_Z$ with increasing redshift.}
    \label{fig:SFE}
\end{figure*}

Here we present the redshift evolution of the star formation efficiency (defined here as $\rm SFE = \int \rm SFR \  dt /\int \dot{M}_{\rm in} \ dt$) in our simulations integrated over full burst cycles.  We remove measurements from burst cycles that are cut off by the final snapshot of the simulations or that have no star formation (due to the tracking of that galaxy being cut off before any star formation as a result of the galaxy merging with a larger galaxy).  The removal of these measurements has no qualitative impact on the results presented in this section but helps isolate the redshift dependence of star formation efficiency within isolated burst cycles.  Figure \ref{fig:SFE} shows that the stellar mass formed per inflowing gas mass increases with increasing redshift for $z=5-12$.  While we do not investigate the physical origin of this evolution in $\rm SFE$ in this work, a plausible explanation is that the denser ISMs of higher-redshift galaxies produce more stars prior to star formation being disrupted by stellar feedback. This evolution is closely tied to the evolution in the metal production efficiency $\varepsilon_Z$ (see Figure \ref{fig:Term_Analysis}), since metal production is closely tied to star formation on short timescales, where core-collapse supernovae and O/B winds are the dominant enrichment channels.

If this evolution trend continues to lower redshift ($z < 5$), the decreased star formation could result in a weakening of the bursty feedback that follows.  The weakened feedback may eventually enable galaxies to retain large amounts of gas between burst cycles, preventing the application of our “Reduced Burst Model" at lower redshift, since galactic evolution could no longer be divided into distinct burst cycles (see the discussion in \S \ref{sec:lowredshift}).

\section{Scaling Relations Driving the Slope of the MZR} \label{appendix:Scaling Relations}

\begin{figure*}[t!]
    \centering
    \includegraphics[width=\linewidth]{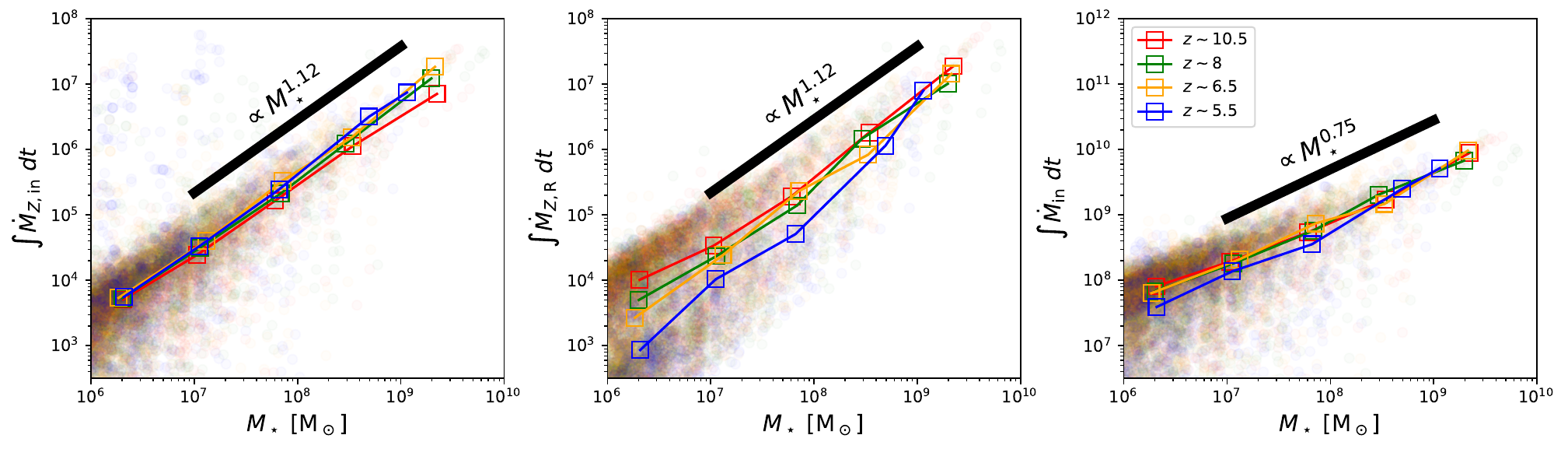}
    \caption{The scaling relations of $\dot{M}_{\rm Z, in}$ (left), $\dot{M}_{\rm Z, R}$ (middle), and $\dot{M}_{\rm in}$ (right) with stellar mass at $z \sim 5.5$ (blue), $6.5$ (orange), $8.0$ (green), and $10.5$ (red).  Smaller, transparent points represent the measurements of individual galaxies at each snapshot within burst cycles.  Empty Squares represent stellar-mass-binned median values.  Solid black lines are meant to guide the eye toward the approximate slope of each scaling relation.  The slope values for these guidelines ($\int\dot{M}_{\rm Z, in}dt,\int\dot{M}_{\rm Z, R}dt \propto M_\star^{1.12}$ and $\int\dot{M}_{\rm in}dt \propto M_\star^{0.75}$) are chosen to demonstrate the ability of these scaling relations within our “Reduced Burst Model" to reproduce the FIRE-2 MZR slope at high redshift measured by \citet{Marszewski_2024}: $Z_{\rm gas} \propto M_\star^{1.12} / M_\star^{0.75} = M_\star^{0.37}$.}
    \label{fig:Scaling_Relations}
\end{figure*}

Figure \ref{fig:Scaling_Relations} shows the scaling relations of each term in our ``Reduced Burst Model", 
\begin{equation} \label{eqn:3rd_simp}
    Z_{\rm gas} = \frac{M_{Z, \rm gas}}{M_{\rm gas}} \approx \frac{\int (\dot{M}_{Z,\rm in} + \dot{M}_{Z,\rm R}) dt}{\int \dot{M}_{\rm in} \ dt},
\end{equation}
with stellar mass throughout our redshift range.  The positive slope of the MZR originates from the numerator of this expression scaling more strongly with stellar mass than the denominator. The scalings summarized in the figure caption quantitatively explain the net MZR slope predicted using FIRE-2 simulations by \cite{Marszewski_2024}.

\section{Burst Cycle and Evacuation Timescales} \label{appendix:Time Scales}

\begin{figure*}[t!]
    \centering
    \includegraphics[width=\linewidth]{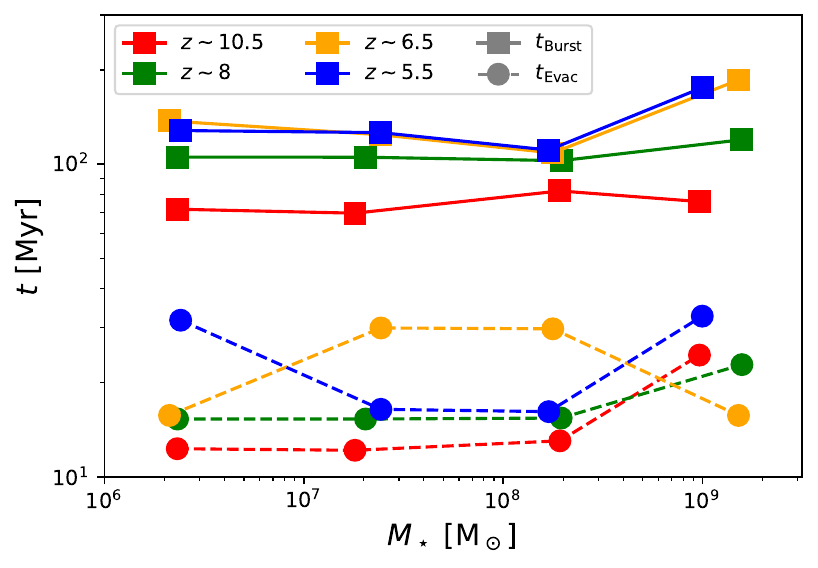}
    \caption{The stellar-mass-binned median burst cycle (squares with solid lines) and evacuation (circles with dashed lines) time scales for galaxies in our sample at $z \sim 5.5$ (blue), $6.5$ (orange), $8.0$ (green), and $10.5$ (red).  Both timescales are weakly dependent on stellar mass but show mild redshift evolution, typically increasing with cosmic time.  The timescale for ISM evacuation ($t_{\rm Evac} \sim 10-30 \ \rm Myr$) is much shorter than the timescale of complete burst cycles ($t_{\rm Burst} \sim 70-200 \ \rm Myr$).}
    \label{fig:Timescales}
\end{figure*}

Here we present the typical timescales over which galactic burst cycles and ISM evacuation occur as a function of stellar mass and redshift.  Burst cycle lengths are calculated as the difference in time between the last and the first snapshot of each burst cycle.  ISM evacuation times are calculated as the difference in time between the last snapshot of each burst cycle and the last snapshot within the burst cycle for which the galaxy contains at least 50 percent of its peak gas mass.  This choice prevents us from misidentifying a minor loss in gas mass (due to, e.g., a minor outflow or astration) as the start of evacuation while also ensuring that a majority of a galaxy's gas mass is evacuated during our defined evacuation time.  For this analysis, we remove measurements from burst cycles that are cut off by the final snapshot of the simulations.  Figure \ref{fig:Timescales} presents the median burst cycle length and ISM evacuation time in different stellar mass and redshift bins.  Note that, using this method, the limited time resolution of our snapshots ($\Delta t_{\rm snapshot} \sim 15 \ \rm Myr$) will generally cause us to overestimate the evacuation time, since evacuation completes between two snapshots but is not resolved until the later snapshot.  The typical magnitude of this overestimation is approximately $\Delta t_{\rm snapshot} / 2$.

Importantly, the timescale over which ISM evacuation occurs ($t_{\rm Evac} \sim 10-30 \ \rm Myr$) is much shorter than the full burst cycle timescale ($t_{\rm Burst} \sim 70-200 \ \rm Myr$) with a typical ratio of $t_{\rm Burst}/t_{\rm Evac} \sim 6$.  This minimizes the amount of time for which the outflow terms in the full “Gas-Regulator Model" (equation \ref{eqn:full_eqn}) can affect a galaxy's metallicity.  This helps justify the omission of these terms from our “Reduced Burst Model" as discussed in Section \ref{sec:develop_analytic_model}. 

Notably, the timescale over which evacuation occurs is longer than the $\rm SFR$ timescale traced by $\rm H\alpha$ emission but shorter than that traced by UV-continuum emission.  As discussed in Section \ref{sec:FMR}, this explains why we find evidence for an FMR-like relationship when using $\rm SFR$ derived from $\rm H\alpha$ emission but not when using $\rm SFR$ derived from UV emission.

\end{appendices}

\end{document}